\documentclass[prd,aps,groupedaddress,showpacs,nobibnotes]{revtex4}
\usepackage{epsf,epsfig,graphicx}
\usepackage{graphics}

\newcounter{muni}

\begin{document}
\hbadness=10000 \pagenumbering{arabic}

\title{Glauber gluons in annihilation amplitudes for heavy meson decays}

\author{Hsiang-nan Li$^{1}$}
\email{hnli@phys.sinica.edu.tw}
\affiliation{$^{1}$Institute of Physics, Academia Sinica, Taipei,
Taiwan 115, Republic of China,}

\date{\today}

\begin{abstract}

We investigate the Glauber divergences in nonfactorizable annihilation amplitudes for
two-body hadronic heavy meson decays in the $k_T$ factorization theorem at one-loop level. 
These divergences can be absorbed into the Glauber factors in the dominant kinematic regions of 
small parton momenta, which modify the interference between a nonfactorizable 
annihilation amplitude and other amplitudes by rotating it with a phase. We postulate that only 
the Glauber effect associated with a pion is significant, due to its special role as a $q\bar q$
bound state and as a pseudo Nambu-Goldstone boson simultaneously. It is elaborated that the data 
of the $D\to\pi\pi$ and $\pi K$ branching ratios have revealed prominent Glauber effects. This work 
provides a solid theoretical ground for the factorization-assisted topological-amplitude parametrization 
of two-body hadronic $D$ meson decays.

\end{abstract}


\maketitle

\section{INTRODUCTION}

The existence, factorization, and impact of a special type of infrared
divergences, called Glauber gluons~\cite{CQ06,Liu:2008cc,Bauer:2010cc,Fleming:2014rea,Gaunt:2014ska,Diehl:2015bca}, 
on studies of two-body hadronic heavy meson decays have been explored thoroughly~\cite{LM11}. 
It was pointed out that the puzzles from the $B\to\pi\pi$ and $\pi\rho$ branching 
ratios, and from the $B\to\pi K$ direct CP asymmetries are attributed to the 
color-suppressed tree amplitudes~\cite{Charng2,Pham:2009ti,CC09}.
We then analyzed radiative corrections to the spectator diagrams for these amplitudes
involved in two-body hadronic $B$ meson decays $B\to M_1M_2$~\cite{LM11}, 
where $M_2$ denotes the meson emitted at the weak vertex. It was found that the Glauber 
divergences are produced in the $k_T$ factorization theorem~\cite{KLS,LUY},
and can be absorbed into additional nonperturbative factors in the dominant kinematic 
regions with small parton momenta~\cite{CL09}. The all-order organization of 
the Glauber divergences follows the standard factorization procedure~\cite{NL03}, 
which relies on the eikonal approximation for soft gluons. The resultant Glauber phase 
factor $\exp(-iS_{e1})$ associated with the $M_1$ meson is the same
for the two leading-order (LO) spectator diagrams~\cite{Li:2014haa}. The Glauber
factors from $M_2$ carry opposite phases, namely, $\exp(iS_{e2})$
for one diagram, and $\exp(-iS_{e2})$ for another~\cite{LM11}. Therefore, they
have different impacts on a spectator amplitude: the latter enhances the
spectator contribution by modifying the interference pattern
between the two LO diagrams. The former rotates
the enhanced spectator contribution by a phase, and changes its
interference with other amplitudes in two-body hadronic $B$ meson decays. 

We postulated that only the Glauber factors associated with a pion give significant 
effects, due to its special role as a $q\bar q$ bound state and as a pseudo 
Nambu-Goldstone (NG) boson simultaneously \cite{NS08}. The Glauber factors 
$\exp(-iS_{e1})$ for $M_1$ and $\exp(\pm iS_{e2})$ for $M_2$ were introduced into 
the spectator amplitudes and treated as additional inputs 
\cite{Liu:2015sra,Liu:2015upa} in the PQCD approach based on the 
$k_T$ factorization theorem. Though a Glauber factor is universal, it causes different 
effects through its convolution with various transverse-momentum-dependent (TMD) 
meson wave functions. It turned out that the Glauber effects from a pion are indeed 
stronger, and improve the consistency between PQCD predictions and experimental data for 
all the $B\to\pi M$ decays with $M=\pi$, $\rho$ and $K$ \cite{Liu:2015sra,Liu:2015upa} . 
In particular, the rotation of the spectator amplitude by $\exp(-iS_{e1})$ is crucial
for enhancing the ratio of the $B^+\to\pi^+\pi^0$ branching fraction over
the $B^0\to\pi^+\pi^-$ one: this ratio depends on
both the color-allowed tree amplitude and the
color-suppressed tree amplitude, so the relative phase between
them matters. It is a nontrivial success that all the puzzles in the $B$ meson decays
mentioned before were resolved at the same time by introducing two Glauber phases.

The dramatic distinction between the measured $D^0\to\pi^+\pi^-$ and
$D^0\to K^+K^-$ branching ratios represented another salient puzzle from $D$ meson decays:
the former (latter) is lower (higher) than theoretical predictions
from analyses based on the topological-amplitude parametrization~\cite{BR10,diag}. 
The deviation in these two modes stands even after flavor SU(3) symmetry breaking 
effects in emission amplitudes were taken into account~\cite{diag}. This subject was 
investigated in the factorization-assisted topological-amplitude (FAT) approach~\cite{Li:2012cfa},
where the Glauber phase was introduced into the nonfactorizable annihilation amplitudes
for two-body hadronic $D$ meson decays. This additional phase modifies 
the interference between the annihilation and emission amplitudes 
involving pions, and improves the overall agreement with data by decreasing the predicted 
$D^0\to\pi^+\pi^-$ branching ratio. Once the Galuber phase
was fixed in a global fit to measured branching ratios in the FAT framework, the
penguin amplitudes, expressed as the combination of the determined hadronic
parameters and the corresponding Wilson coefficients, were also obtained
accordingly. We then predicted direct CP asymmetries
in $D\to PP$ decays, with $P$ denoting a pseudoscalar meson, in the Standard Model without ambiguity. 
Especially, we predicted the difference between the two direct CP asymmetries,
$\Delta A_{\rm CP}\equiv A_{\rm CP}(D^0\to K^+K^-)-A_{\rm CP}(D^0\to\pi^+\pi^-)
\approx -1.0\times 10^{-3}$~\cite{Li:2012cfa}, which was verified by the more precise 
LHCb data~\cite{Aaij:2014gsa,Aaij:2016cfh} announced later. We mention that a prediction for 
$\Delta A_{\rm CP}$ similar to ours was made in the topological-amplitude
approach combined with final-state rescattering~\cite{Cheng:2012wr}.

However, the existence and factorization of the Glauber divergences in the 
nonfactorizable annihilation diagrams were not verified rigorously in Ref.~\cite{Li:2012cfa}.
Here we will explore the Glauber gluons in these diagrams in the $k_T$ factorization
theorem, and show that they can be factorized into the nonperturbative
phase factors associated with final-state mesons in the kinematic regions with
small parton momenta. It is found that the Glauber phases are the same for 
the two LO nonfactorizable annihilation diagrams, so the Glauber effect rotates 
a nonfactorizable annihilation amplitude, and modifies its interference with other amplitudes.
To reveal the Glauber effects in two-body hadronic $D$ meson decays, we highlight several 
$D\to\pi\pi$ and $\pi K$ modes, whose measured branching ratios demand the inclusion of 
the Glauber phases in the FAT approach. This work provides a solid theoretical ground
for the FAT parametrization, and confirms the speculation 
that the Glauber effects associated with a pion are crucial for resolving the puzzles 
in heavy meson decays \cite{Zhou:2015jba}.

In Sec.~II we analyze the collinear divergences in radiative corrections to
the nonfactorizable annihilation amplitudes in the $k_T$ factorization theorem at one-loop
level, and identify the residual infrared divergences caused by the Glauber gluons that
cannot be absorbed into TMD meson wave functions. 
The Glauber divergences are factorized out of the two LO nonfactorizable annihilation diagrams
and grouped into the nonperturbative phase factors. In Sec.~III we discuss
the data for the $D\to\pi\pi$ and $D\to\pi K$ branching ratios, which exhibit the 
apparent Glauger effects. Section IV contains the conclusion.

\section{FACTORIZATION OF GLAUBER GLUONS}

\begin{figure}[t]
\begin{center}
\includegraphics[scale=1.1]{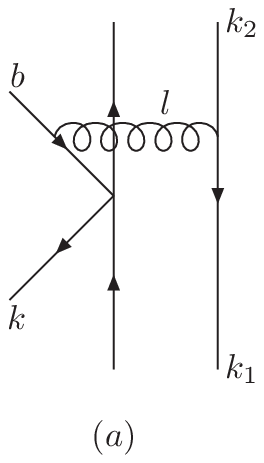}\hspace{1cm}
\includegraphics[scale=1.1]{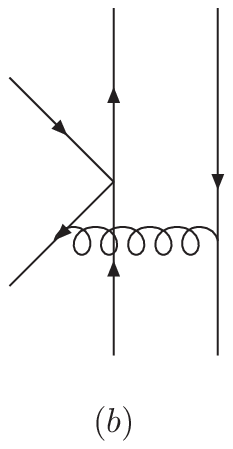}
\caption{LO diagrams for a nonfactorizable annihilation amplitude.} \label{fig1}
\end{center}
\end{figure}

We examine the infrared divergences in the next-to-leading-order (NLO) nonfactorizable 
annihilation diagrams, following the reasoning in Ref.~\cite{LM11}.
Consider the $B(P_B)\to M_1(P_1)M_2(P_2)$ decay, where $P_B$, $P_1$,
and $P_2$ are the momenta of the $B$, $M_1$, and $M_2$ mesons,
respectively. For convenience, we choose $P_B=(P_B^+,P_B^-,{\bf
0}_T)$ with $P_B^+=P_B^-=m_B/\sqrt{2}$, $m_B$ being the $B$ meson
mass, and $P_1$ ($P_2$) in the plus (minus) direction. The parton
momenta $k$, $k_1$ and $k_2$ for the $B$, $M_1$ and $M_2$ mesons, respectively,
are labelled in Fig.~\ref{fig1}(a), which are assumed to obey the hierarchy in the regions
of small parton momenta~\cite{LSW12},
\begin{eqnarray}
m_B \gg k_1^+, k_2^- \gg k^\mu, k_{iT}\sim O(\Lambda),\label{hie}
\end{eqnarray}
with $i=1,2$ and $\Lambda$ representing a small scale. The even smaller components $k_1^-$ and
$k_2^+$ under the above hierarchy have been neglected.

\subsection{NLO Corrections to Fig.~\ref{fig1}(a)}

\begin{figure}[t]
\begin{center}
\includegraphics[scale=1.1]{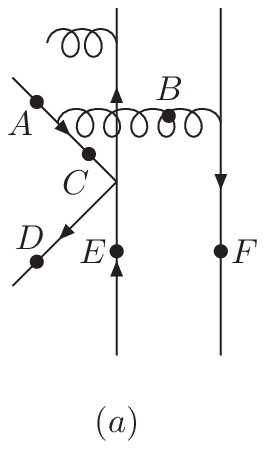}\hspace{1cm}
\includegraphics[scale=1.1]{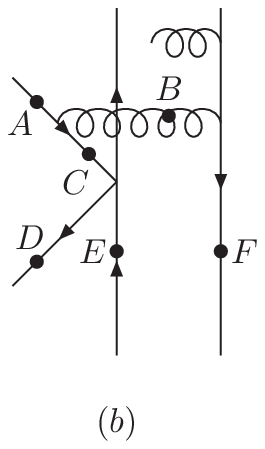}
\caption{NLO diagrams for Fig.~\ref{fig1}(a), where the radiative gluon is emitted
by (a) the valence quark and (b) the valence anti-quark of the $M_2$ meson, and 
attaches to the vertices labelled by $A$, $B$,..., or $F$.} \label{fig2}
\end{center}
\end{figure}

We first search for the Glauber gluons associated with the LO nonfactorizable annihilation
diagram in Fig.~\ref{fig1}(a), 
starting with the set of NLO diagrams in Fig.~\ref{fig2}.
Due to the soft cancellation between the gluons radiated by the valence quark
and by the valence anti-quark of the $M_2$ meson~\cite{LT98}, only the collinear
region with the loop momentum $l$ being collimated to $P_2$ is relevant, and the particles in 
the $B$ and $M_1$ mesons, to which the collinear gluons attach, are off-shell.
The transverse-momentum dependence of these propagators is thus negligible, which
can then be approximated by the eikonal propagators
$1/(l^-\pm i\epsilon)$. For a loop diagram to generate an imaginary
Glauber logarithm, a necessary (but not sufficient) condition is
that the interval of $l^-$ covers the origin $l^-=0$. The
corresponding integrals then contain imaginary pieces,
\begin{eqnarray}
{\rm Im} \int \frac{dl^-}{l^- \pm i\epsilon}
=\mp\pi \int dl^- \delta(l^-)=\mp\pi,\label{int}
\end{eqnarray}
under the principal-value prescription. As the radiative gluon goes on shell,
an imaginary logarithm appears. It is known that a TMD meson wave function 
does not involve imaginary infrared logarithms. Hence, the Glauber logarithm is 
residual, and cannot be absorbed into the $M_2$ meson wave function in view of its universality.

Start with Fig.~\ref{fig2}(a), where the radiative gluon is
emitted by the valence quark of the $M_2$ meson and attaches to the other lines. The reducible
diagrams, such as the self-energy correction to the valence quark of $M_2$, are not displayed.
For the attachments $A$, $C$, $D$ and $E$, the pole structures of the loop integrands on the $l^+$ 
plane imply that only the $l^-<0$ range contributes to the contour integrals over $l^+$. Taking the 
attachment $C$ as an example, we have the integrand proportional to
\begin{eqnarray}
\frac{1}{2(l^-+{\bar k}_2^-)l^+-|{\bf k}_{2T}-{\bf l}_T|^2+i\epsilon}
\frac{1}{2l^-l^+-l_T^2+i\epsilon}\frac{1}{2(l^-+{\bar k}_2^-)(l^++{\bar k}_1^+)-m_b^2
+i\epsilon}, \label{ac}
\end{eqnarray}
where ${\bar k}_1\equiv P_1-k_1$, ${\bar k}_2=P_2-k_2$, and the $k^\pm$ and TMD terms in the hard $b$ 
quark propagator have been dropped based on Eq.~(\ref{hie}) for simplicity. When poles are located 
in the different half planes of $l^+$, the loop integral does not vanish, no matter whether the contour 
of $l^+$ is closed from the upper or lower half plane. It is the case for Eq.~(\ref{ac}) only when $l^-<0$ 
obviously. Then the attachment $C$ does not contribute a Glauber divergence according to Eq.~(\ref{int}).
A general observation is that a simple diagram like a vertex correction, ie., the attachment $C$, $D$ 
or $E$, does not develop a Glauber divergence.

The attachment $B$, with the radiative gluon attaching to a hard gluon,
does not produce a Glauber divergence either. The corresponding integrand proportional to
\begin{eqnarray}
& &\frac{1}{2(l^-+{\bar k}_2^-)l^+-|{\bf k}_{2T}-{\bf l}_T|^2+i\epsilon}
\frac{1}{2l^-l^+-l_T^2+i\epsilon}
\frac{1}{2(l^--k_2^-)(l^+-k_1^+)-|{\bf k}_{1T}+{\bf k}_{2T}-{\bf l}_T|^2+i\epsilon}\nonumber\\
& &\times \frac{1}{2(l^-+{\bar k}_2^-)(l^++{\bar k}_1^+)-m_b^2+i\epsilon},
\end{eqnarray}  
indicates that the loop integral does not vanish as
$-{\bar k}_2^-<l^-<k_2^-$. We focus on the poles in the lower half plane of $l^+$.
The pole $l^+=l_T^2/(2l^-)$ of the second, ie., radiative gluon propagator moves from the
lower half plane to the upper one, when $l^-$ varies from $l^->0$ to $l^-<0$. That is, its
residue does not exist in both the positive and negative regions of $l^-$ required by Eq.~(\ref{int}). 
The pole $l^+=m_b^2/[2(l^-+{\bar k}_2^-)]-{\bar k}_1^+$ of the last, ie., $b$ quark propagator
yields an off-shell quark in the $M_2$ meson denoted by the first propagator. 
Namely, it does not contribute a collinear configuration associated with $M_2$. 
We thus pick up the pole $l^+=|{\bf k}_{2T}-{\bf l}_T|^2/[2(l^-+{\bar k}_2^-)]\sim
O(\Lambda^2/m_B)$ from the first propagator, which is much smaller than $m_B$ and 
makes a collinear configuration. However, the denominators for both the hard gluon corresponding 
to the third propagator and the hard $b$ quark do not flip sign in the range $-{\bar k}_2^-<l^-<k_2^-$, 
and thus no imaginary pieces are generated.

We conclude that the attachments $A$-$E$ in Fig.~\ref{fig2}(a) contribute only to the
$M_2$ meson wave function, whose collinear divergences can be collected by the Wilson links
resulting from the eikonal approximation. The detailed procedure for factorizing 
these NLO diagrams is referred to \cite{NL03}.

Possible Glauber divergences may appear in the attachment $F$, whose integrand contains
the denominator
\begin{eqnarray}
[({\bar k}_2+l)^2+i\epsilon][({\bar k}_1+{\bar k}_2-k+l)^2-m_b^2+i\epsilon]
(l^2+i\epsilon)[(k_1-l)^2+i\epsilon][(k_1+k_2-l)^2+i\epsilon].\label{pf}
\end{eqnarray}
Nonvanishing contributions come from the range $0< l^- < k_2^-$
($-{\bar k}_2^-+k^- < l^- <0$, $-{\bar k}_2^-<l^-<-{\bar k}_2^-+k^- $), where
the poles of $l^+$ are given by
\begin{eqnarray}
& &l^+=\frac{|{\bf l}_T-{\bf k}_{2T}|^2}{2(l^-+{\bar k}_2^-)}-i\epsilon(-i\epsilon,\;-i\epsilon),\label{mo1}\\
& &l^+=-{\bar k}_1^++k^++\frac{m_b^2
}{2(l^-+{\bar k}_2^--k^-)}
-i\epsilon(-i\epsilon,\;+i\epsilon),\label{mo5}\\
& &l^+=\frac{l_T^2}{2l^-}-i\epsilon (+i\epsilon,\; +i\epsilon),\label{mo4}\\
& &l^+=k_1^++\frac{|{\bf l}_T-{\bf k}_{1T}|^2}{2l^-}-i\epsilon(+i\epsilon,\;+i\epsilon),\label{mo2}\\
& &l^+=k_1^++\frac{|{\bf l}_T-{\bf k}_{1T}-{\bf k}_{2T}|^2}{2(l^--k_2^-)}
+i\epsilon(+i\epsilon,\;+i\epsilon).\label{mo3}
\end{eqnarray}
We pick up the poles in the lower half plane of $l^+$. 
The first pole $l^+\sim O(\Lambda^2/m_B)$ contributes to the loop
integral in the range $-{\bar k}_2^-<l^-<k_2^-$, which covers the origin $l^-=0$,
and also makes a collinear configuration associated with 
the $M_2$ meson. It is then seen that the loop integral develops a Glauber divergence 
from the eikonalized spectator propagator $1/(k_1-l)^2$ and the on-shell radiative gluon. 
The second pole $l^+\sim O(k_1^+)$ in Eq.~(\ref{mo5}), contributing to the loop
integral in the range $-{\bar k}_2^-+k^-<l^-<k_2^-$, does not correspond to the considered 
collinear configuration.

An alternative way to verify the existence of the Glauber divergence  in the attachment $F$ follows the derivation
in Ref.~\cite{CQ06}: we eikonalize the quark line with the momentum $k_1-l$ first, and focus 
on the imaginary piece proportional to $\delta(l^-)$. The integration over $l^-$ leads the 
radiative gluon propagator to $1/l_T^2$, and the quark propagator on the $M_2$ side and the
hard gluon propagator to
\begin{eqnarray} 
\frac{1}{2{\bar k}_2^-l^+-|{\bf k}_{2T}-{\bf l}_T|^2+i\epsilon}
\frac{1}{2k_2^-(k_1^+-l^+)-|{\bf k}_{1T}+{\bf k}_{2T}-{\bf l}_T|^2+i\epsilon}.\label{alt}
\end{eqnarray}
We do not show the hard $b$ quark propagator, since its $l$ dependence does not
affect the conclusion. The above expression implies clearly that the contour integral
over $l^+$ does not vanish, because the two poles of $l^+$ are located on different 
half planes, and that the Glauber divergence from $l_T\to 0$ stands. 
It is also easy to explain by means of Eq.~(\ref{alt})
that a Glauber divergence does not appear in the collinear
factorization theorem for two-body hadronic $B$ meson decays \cite{Beneke:1999br},
which assumes the dominance of the region with large parton momenta $k_1^+\sim k_2^-\sim O(m_B)$. 
The small component $l^+$ in the collinear configuration is then negligible in the hard gluon and 
$b$ quark propagators, such that only a single pole of $l^+$ from the first propagator
in Eq.~(\ref{alt}) remains, and the contour integral over $l^+$ vanishes.
In other words, a Glauber divergence is power suppressed in the collinear factorization.

We then study the infrared divergences from Fig.~\ref{fig2}(b) with the radiative gluon being
emitted by the valence anti-quark of the $M_2$ meson and attaching to the other lines. Similarly,
for the attachments $A$, $B$ and $F$, the pole structures of the loop integrands on the $l^+$ plane indicate that
only the $l^-<0$ range contributes to the contour integrals over $l^+$. The radiative gluon attaches 
to the hard $b$ quark in the attachment $C$. As elaborated before, no Glauber divergence exists in the
above cases. 

The integrand for the attachment $D$ has five denominators,
\begin{eqnarray}
[(k_1+k_2+l)^2+i\epsilon][(k_2+l)^2+i\epsilon]
[(k+l)^2+i\epsilon](l^2+i\epsilon)[({\bar k}_1+{\bar k}_2-k-l)^2-m_b^2+i\epsilon].\label{2e}
\end{eqnarray}
Nonvanishing contributions arise from the range $0< l^- < {\bar k}_2^--k^-$
($-k^- < l^- <0$, $-k_2^-<l^-<-k^-$), where
the poles of $l^+$ are given by
\begin{eqnarray}
& &l^+=-k_1^++\frac{|{\bf l}_T+{\bf k}_{1T}+{\bf
k}_{2T}|^2}{2(l^-+k_2^-)}-i\epsilon(-i\epsilon,\;-i\epsilon),\label{po1}\\
& &l^+= \frac{|{\bf l}_T+{\bf
k}_{2T}|^2}{2(l^-+k_2^-)}-i\epsilon(-i\epsilon,\;-i\epsilon),\label{po2}\\
& &l^+=-k^++\frac{|{\bf l}_T+{\bf k}_T|^2}{2(l^-+k^-)}
-i\epsilon(-i\epsilon,\;+i\epsilon),\label{po3}\\
& &l^+=\frac{l_T^2}{2l^-}-i\epsilon (+i\epsilon,\; +i\epsilon),\label{po4}\\
& &l^+={\bar k}_1^+-k^++\frac{m_b^2}{2(l^--{\bar k}_2^-+k^-)}
+i\epsilon(+i\epsilon,\;+i\epsilon).\label{po5}
\end{eqnarray}
It is more convenient to pick up the poles in the upper half
plane of $l^+$, and noticed that only the last pole contributes in the range $-k_2^-<l^-<{\bar k}_2^--k^-$
covering the origin $l^-=0$. However, this pole sets a large
$l^+$ component, with which the other four propagators, being off-shell, do not generate Glauber 
divergences. 

The poles involved in the attachment $E$ are similar to those in the attachment $D$,
but with the third one in Eq.~(\ref{po3}) being replaced by
\begin{eqnarray}
l^+={\bar k}_1^++\frac{|{\bf l}_T+{\bf k}_{1T}|^2}{2l^-}
-i\epsilon(+i\epsilon,\;+i\epsilon).\label{no3}
\end{eqnarray}
The same argument applies to this case apparently, and we conclude
that the attachment $E$ does not yield a Glauber divergence either. 
That is, the diagrams in Fig.~\ref{fig2}(b) contribute only to the
$M_2$ meson wave function, whose collinear divergences can be collected by the Wilson links.

The eikonalization of the propagator $1/[(k_1-l)^2+i\epsilon]$ for the attachment $F$ in 
Fig.~\ref{fig2}(a) leads to $1/(-l^-+i\epsilon)$ \cite{LM11}, which contributes an imaginary
piece $-\pi i \delta(l^-)$. The principal value, ie., the real piece of this eikonal propagator
gives rise to a Wilson link, which goes into the definition of the $M_2$ meson wave 
function \cite{NL03}. The NLO residual infrared divergence is then written as
\begin{eqnarray}
& &-\frac{g^2}{2N_c}\int \frac{d^4l}{(2\pi)^4}tr\bigg[...\gamma_5\not\!\! P_2(-ig\gamma^-)\frac{i(\not\!
{\bar k}_2+\not l)}{(k_2+l)^2+i\epsilon}\bigg]\nonumber\\
& &\times\frac{-i}{(k_1+k_2-l)^2+i\epsilon}
\frac{-i}{l^2+i\epsilon}(-\pi i) \delta(l^-),\label{il}
\end{eqnarray}
where $...$ denotes the rest of the integrand, and $\gamma_5\!\not\!\!P_2$ comes from the 
twist-2 structure of the $M_2$ meson wave function. Note that this Glauber divergence carries
the color factor $-1/(2N_c)$ \cite{LT98}, $N_c$ being the number of colors.
Because the pole  $l^+\sim O(k_1^+)$ in Eq.~(\ref{mo3}) in the upper half
plane stays far away from the selected pole $l^+\sim O(\Lambda^2/m_B)$, we can 
always deform the contour of $l^+$, such that $l^+$ remains at least $O(\Lambda)$,
and the hierarchy
\begin{eqnarray}
{\bar k}_2^-l^+ \sim O(m_B\Lambda)\gg |{\bf l}_T-{\bf k}_{2T}|^2
\sim O(\Lambda^2),
\end{eqnarray}
holds. The quark with the momentum ${\bar k}_2+l$ in
Eq.~(\ref{il}) can then be eikonalized into $1/(l^++i\epsilon)$.
The above argument goes exactly like that applied to the factorization of the 
Glauber gluons from the low-$p_T$ hadron hadroproduction \cite{CL09}.

Equation~(\ref{il}) is thus factorized into
\begin{eqnarray}
& &\frac{g^2}{2N_c}\int\frac{d^4l}{(2\pi)^4}tr\bigg[...\gamma_5\not\!\! P_2
\bigg]\frac{-i}{(k_1+k_2+l)^2+i\epsilon}\nonumber\\ &
&\times\frac{1}{l^++i\epsilon}\frac{-i}{l^2+i\epsilon}\pi i
\delta(l^-).\label{il2}
\end{eqnarray}
We close the contour in the lower half plane of $l^+$, pick up
the pole $l^+= 0-i\epsilon$ from the eikonal propagator $1/(l^++i\epsilon)$, 
and derive the convolution in the transverse momentum of the Glauber gluon, 
\begin{eqnarray}
\frac{i}{2N_c}\frac{\alpha_s}{2\pi}
\int\frac{d^2l_T}{l_T^2}{\cal
M}_a^{(0)}({\bf l}_T). \label{vi2}
\end{eqnarray}
The imaginary logarithm is explicit in the above expression, and
${\cal M}_a^{(0)}$ denotes the LO nonfactorizable annihilation amplitude from Fig.~\ref{fig1}(a).

\subsection{NLO Corrections to Fig.~\ref{fig1}(b)}

\begin{figure}[t]
\begin{center}
\includegraphics[scale=1.1]{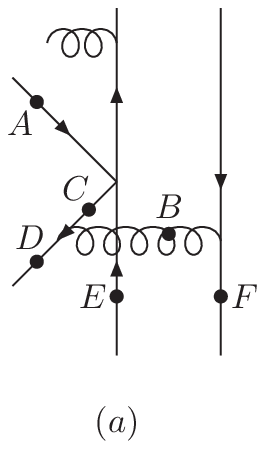}\hspace{1cm}
\includegraphics[scale=1.1]{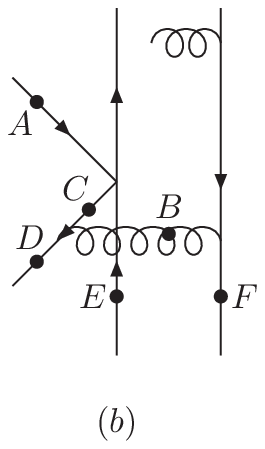}
\caption{NLO diagrams for Fig.~\ref{fig1}(b), where the radiative gluon is emitted
by (a) the valence quark and (b) the valence anti-quark of the $M_2$ meson, and attaches 
to the vertices labelled by $A$, $B$,..., or $F$.} \label{fig3}
\end{center}
\end{figure}

We search for the Glauber divergences in the NLO
corrections to Fig.~\ref{fig1}(b), which include the attachments of the radiative gluon
emitted by the valence quark of the $M_2$ meson as depicted in Fig.~\ref{fig3}(a).
The attachments $A$, $C$ and $E$, being the simple vertex corrections without
Glauber divergences, contribute only to the $M_2$ meson wave function.
The attachment $B$ on the hard gluon does not produce
a Glauber divergence. We then discuss the attachment $D$, whose integrand  
is proportional to
\begin{eqnarray}
& &\frac{1}{2(l^-+{\bar k}_2^-)l^+-|{\bf k}_{2T}-{\bf l}_T|^2+i\epsilon}
\frac{1}{2l^-l^+-l_T^2+i\epsilon}
\frac{1}{2(l^-+k^-)(l^++k^+)-|{\bf k}_{T}+{\bf l}_T|^2+i\epsilon}\nonumber\\
& &\times \frac{1}{2(l^-+k^--k_2^-)(l^++k^+-k_1^+)
-|{\bf k}_{T}-{\bf k}_{1T}-{\bf k}_{2T}+{\bf l}_T|^2+i\epsilon}.
\end{eqnarray} 
Similar to the attachment $D$ in Fig.~\ref{fig2}(b), it is more convenient to pick up the poles in the upper half
plane of $l^+$, and noticed that only the last pole contributes in the range $-{\bar k}_2^-<l^-<k_2^--k^-$
covering the origin $l^-=0$. However, this pole specifies a large
$l^+\approx k_1^+$ component, with which the other three propagators stay off-shell, and 
do not generate Glauber divergences.  

The integrand for the attachment $F$ in Fig.~\ref{fig3}(a) has the denominator
\begin{eqnarray}
[({\bar k}_2+l)^2+i\epsilon](l^2+i\epsilon)[(k_1-l)^2+i\epsilon][(k-k_1-k_2+l)^2+i\epsilon]
[(k_1+k_2-l)^2+i\epsilon].\label{qf}
\end{eqnarray}
Nonvanishing contributions come from the range $k_2^--k^-<l^-<k_2^- $
($0< l^- < k_2^--k^-$, $-{\bar k}_2^- < l^- <0$), where
the poles of $l^+$ are given by
\begin{eqnarray}
& &l^+=\frac{|{\bf l}_T-{\bf k}_{2T}|^2}{2(l^-+{\bar k}_2^-)}-i\epsilon(-i\epsilon,\;-i\epsilon),\label{qo1}\\
& &l^+=\frac{l_T^2}{2l^-}-i\epsilon (-i\epsilon,\; +i\epsilon),\label{qo4}\\
& &l^+=k_1^++\frac{|{\bf l}_T-{\bf k}_{1T}|^2}{2l^-}-i\epsilon(-i\epsilon,\;+i\epsilon),\label{qo2}\\
& &l^+=k_1^+-k^++\frac{|{\bf l}_T-{\bf k}_{1T}-{\bf
k}_{2T}+{\bf k}_T|^2}{2(l^--k_2^-+k^-)}
-i\epsilon(+i\epsilon,\;+i\epsilon),\label{qo5}\\
& &l^+=k_1^++\frac{|{\bf l}_T-{\bf k}_{1T}-{\bf k}_{2T}|^2}{2(l^--k_2^-)}
+i\epsilon(+i\epsilon,\;+i\epsilon).\label{qo3}
\end{eqnarray}
We pick up the poles in the lower half plane of $l^+$, and find that 
only the first pole $l^+\sim O(\Lambda_{\rm QCD}^2/m_B)$ contributes to the loop
integral in the range $-{\bar k}_2^- < l^- <k_2^-$ covering the origin $l^-=0$.
It is also the pole that corresponds to the considered collinear configuration associated with 
the $M_2$ meson. It is seen that the loop integral develops a Glauber divergence from the 
eikonalized quark propagator $1/(k_1-l)^2$ and the on-shell radiative gluon, 
with the color factor the same as of the attachment $F$ in Fig.~\ref{fig2}(a).

At last, we examine the infrared divergences from Fig.~\ref{fig3}(b) with the radiative gluon being
emitted by the valence anti-quark of the $M_2$ meson and attaching to the other lines.   
The integrands for the attachments $A$, $D$ and $E$ contain the denominators
\begin{eqnarray}
& &[(k_1+k_2+l)^2+i\epsilon][(k_2+l)^2+i\epsilon]
(l^2+i\epsilon)[(P_B-k+l)^2-m_b^2+i\epsilon][(k-k_1-k_2-l)^2+i\epsilon],\label{3a}\\
& &[(k_1+k_2+l)^2+i\epsilon][(k_2+l)^2+i\epsilon]
[(k+l)^2+i\epsilon](l^2+i\epsilon),\label{3d}\\
& &[(k_1+k_2+l)^2+i\epsilon][(k_2+l)^2+i\epsilon](l^2+i\epsilon)
[({\bar k}_1-l)^2+i\epsilon][(k-k_1-k_2-l)^2+i\epsilon],\label{3e}
\end{eqnarray}
respectively. It is easy to show that these attachments do not produce 
Glauber divergences, simply because only the range $l^-<0$ contributes to the loop integrals.
The attachments $B$ and $C$, located on the virtual lines, do not
either. The attachment $F$, representing a vertex correction,
is free of a Glauber divergence. That is, all the attachments in Fig.~\ref{fig3}(b)
contribute only to the $M_2$ meson wave function.

The NLO residual infrared divergence in the attachment $F$ in Fig.~\ref{fig3}(a) 
is then extracted from the Glauber region and collected by the integral similar to Eq.~(\ref{il}).
The argument leading to Eq.~(\ref{il2}) also applies:
since the poles  $l^+\sim O(k_1^+)$ in Eqs.~(\ref{qo5}) and (\ref{qo3}) in the upper half
plane stay far away from the selected pole $l^+\sim O(\Lambda^2/m_B)$, we can 
always deform the contour of $l^+$, such that the quark carrying the momentum ${\bar k}_2+l$ 
is eikonalized into $1/(l^++i\epsilon)$. We then obtain explicitly the imaginary logarithm:
\begin{eqnarray}
\frac{i}{2N_c}\frac{\alpha_s}{2\pi}
\int\frac{d^2l_T}{l_T^2}{\cal M}_b^{(0)}({\bf l}_T), \label{vi3}
\end{eqnarray}
where ${\cal M}_b^{(0)}$ represents the LO nonfactorizable annihilation amplitude from
Fig.~\ref{fig1}(b). It is noticed that Eq.~(\ref{vi3}) has a sign the same as of
Eq.~(\ref{vi2}). This observation differs from that for the spectator amplitudes made 
in \cite{Li:2014haa}: the Glauber logarithms for the two LO spectator amplitudes 
associated with the $M_2$ meson emitted at the weak vertex are opposite in sign.

The Glauber divergences associated with the $M_1$ meson are
analyzed in the same way, and the results are similar to Eqs.~(\ref{vi2}) and (\ref{vi3}).
In this case a Glauber gluon is emitted by the valence anti-quark in the $M_1$ meson
and attaches to the valence anti-quark in the $M_2$ meson, so the Glauber divergences 
have the same sign as Eqs.~(\ref{vi2}) and (\ref{vi3}).
The exponentiation of the NLO results in Eqs.~(\ref{vi2}) and (\ref{vi3}) yields 
\begin{eqnarray}
M_a^G &=& \exp(iS_{G}){\cal M}_a^{(0)},\nonumber\\
M_b^G &=& \exp(iS_{G}){\cal M}_b^{(0)},\label{iab}
\end{eqnarray}
where both the LO amplitudes have been rotated by the same Glauber phase $S_G$.
Equation~(\ref{iab}) concludes our investigation of the Glauber divergences in
the nonfactorizable annihilation diagrams for the two-body hadronic heavy meson decays.
It also holds for nonfactorizable $W$-exchange amplitudes, since 
fermion flows do not affect the derivation of the Glauber factors.
A definition for the Glauber factor in terms of a matrix element of
four Wilson links has been constructed in \cite{CL09}.
Strictly speaking, Eq.~(\ref{iab}) should be understood as convolutions
between the Glauber factors and the LO nonfactorizable annihilation amplitudes in the
impact parameter space \cite{Liu:2015sra}. Nevertheless, treating $S_{G}$ as a constant 
parameter is convenient for phenomenological applications, such as the FAT parametrization 
for two-body hadronic heavy meson decays.

\section{PHENOMENOLOGICAL EFFECTS}

It has been shown that the nonfactorizable annihilation contribution is less than
10\% of the factorizable one in two-body hadronic $B$ meson decays~\cite{Keum:2000ms}. 
The Glauber factor obtained here introduces an overall phase to the nonfactorizale 
annihilation amplitude, instead of changing the interference pattern between the two LO hard
diagrams. This contribution is thus expected to remain negligible, and 
does not have a significant impact on, say, the $B\to \pi\pi$ and $B\to\pi K$ decays. 
That is, the $B\to \pi\pi$ and $B\to\pi K$ puzzles mentioned before are resolved mainly by the Glauber 
effects on the spectator contributions~\cite{LM11,Li:2014haa,Liu:2015sra,Liu:2015upa}.
The Glauber factor in Eq.~(\ref{iab}) may be crucial for a decay dominated by the tree 
annihilation topology, to which the factorizable annihilation contribution is suppressed 
by the helicity conservation. The nonfactorizale and factorizable
annihilation contributions then become comparable, such that the rotation of the 
former can give an effect. The $B_c\to\pi K$, $\pi\eta$ and $\pi\eta'$ decays, belonging 
to this category, are appropriate for probing the Glauber effect discussed in this work. 
However, these modes with small branching ratios about $10^{-7}$-$10^{-8}$~\cite{Liu:2009qa}
are not easily accessed.

$D$ meson decays provide another potential arena for testing the Glauber effect
on the nonfactorizable annihilation contributions.
Penguin amplitudes in these decays are usually tiny, and annihilation
contributions are not really power suppressed compared to emission ones, because the
$D$ meson mass is not much higher than the QCD scale. For the same reason, 
the nonfactorizable annihilation contribution may be comparable to the factorizable one.
The Glauber factor $\exp(iS_G)$ has been
introduced as a free parameter in the FAT approach to two-body hadronic $D$ meson decays,
and associated with a final-state pion in the nonfactorizable annihilation and $W$-exchange channels.
The global fits to the measured $D\to PP$ and $PV$ branching ratios confirmed
that the Glauber phases are substantial, $S_G=-0.50$ \cite{Li:2012cfa} 
and $S_G=-0.85$ with the $\rho$-$\omega$ mixing being taken into account \cite{Li:2013xsa}, respectively. 
The puzzle from the data of the $D^0\to\pi^+\pi^-$ and $D^0\to K^+K^-$ branching ratios
stated in the Introduction was then resolved~\cite{Li:2012cfa}. These sizable Glauber phases from 
the global analyses indicate that they are demanded by modes involving pions in general.

Below we manifest the Glauber effect in two-body hadronic $D$ meson decays from a different 
viewpoint: we identify several specific modes, whose data exhibit the 
impact of the Glauber phase clearly. Consider the doubly-Cabibbo-suppressed 
$D^+\to \pi^0K^+$ and $D^0\to \pi^-K^+$ decays, both of which involve the color-allowed
emission amplitude with the $D\to \pi$ transition. The former (latter) also proceeds via
the annihilation ($W$-exchange) process with the $u\bar u$ 
quark pair popping out of the vacuum. It is expected that the
properties of individual mesons which may break the SU(3) symmetry, such as the 
decay constants, cancel largely in the ratio of their
branching fractions. The corresponding singly-Cabibbo-suppressed 
$D^+\to K_SK^+$ and $D^0\to K^-K^+$ decays involve the color-allowed
emission amplitude with the $D\to K$ transition. The former (latter) also proceeds via
the annihilation ($W$-exchange) process with the $s\bar s$ ($u\bar u$) 
quark pair popping out of the vacuum.  The SU(3) symmetry breaking
effects also cancel largely in the ratio of their branching fractions. Note that the
emission and annihilation amplitudes in the $D^+\to \pi^0K^+$ decay are opposite in sign,
attributed to the $d\bar d$ ($u\bar u$) component of the $\pi^0$ meson contributing to the former
(latter). This opposite sign in the corresponding $D^+\to K_SK^+$ decay is provided
by the different Cabibbo-Kobayashi-Maskawa (CKM) factors associated with the two amplitudes.

The similarity between the above $D\to\pi K$ and $D\to KK$ sets suggests
\begin{eqnarray}
\frac{B(D^+\to \pi^0K^+)}{B(D^0\to \pi^-K^+)}\approx 
\frac{B(D^+\to K_SK^+)}{B(D^0\to K^-K^+)}.\label{r1}
\end{eqnarray}
However, the data of the branching ratios $B(D^+\to \pi^0K^+)=(2.08\pm 0.21)\times 10^{-4}$, 
$B(D^0\to \pi^-K^+)=(1.50\pm 0.07)\times 10^{-4}$, 
$B(D^+\to K_SK^+)=(3.04\pm 0.09)\times 10^{-3}$ and 
$B(D^0\to K^-K^+)=(4.08\pm 0.06)\times 10^{-3}$~\cite{PDG}
lead to $1.39\pm 0.15$ for the left-hand side and $0.745\pm 0.025$
for the right-hand side, which differ significantly. The error for the
ratio is estimated by summing the fractional errors of the numerator and 
the denominator in quadrature. It is very difficult
to accommodate the above data without the Glauber effect associated with a pion.

Another example comes from the comparison of the $D\to \pi\pi$ and $\pi K$ decays.
Both the singly-Cabibbo-suppressed 
$D^0\to \pi^-\pi^+$ and $D^0\to \pi^0\pi^0$ decays proceed via
the $W$-exchange process with the light (non-strange)
quark pairs popping out of the vacuum. The former (latter) also involves the 
color-allowed (color-suppressed) emission amplitude with the $D\to \pi$ transition. 
The corresponding modes are the doubly-Cabibbo-suppressed $D^0\to \pi^-K^+$
decay and the Cabibbo-favored $D^0\to \pi^0K_S$ decay, which involve the color-allowed
and color-suppressed emission amplitudes, respectively, and the $W$-exchange process with 
the light quark pairs popping out of the vacuum. The Cabibbo-favored $D^0\to \pi^+K^-$ 
decay can be also considered, which results from the color-allowed emission amplitude with the 
$D\to K$ transition, and the $W$-exchange channel with 
the light quark pair popping out of the vacuum. The similarity 
among the above sets of modes implies
\begin{eqnarray}
\frac{B(D^0\to \pi^-\pi^+)}{B(D^0\to \pi^0\pi^0)}\approx 
\frac{ B(D^0\to \pi^-K^+)}{\lambda^4B(D^0\to \pi^0K_S)}\approx 
\frac{ B(D^0\to \pi^+K^-)}{B(D^0\to \pi^0K_S)},\label{r2}
\end{eqnarray}
where the denominator of the second ratio has been corrected by the Wolfenstein
parameter $\lambda=0.23$. 
Since the $W$-exchange amplitudes in all the four modes 
arise from the light quark pair production in the vacuum, the first equality 
is expected to hold better than Eq.~(\ref{r1}). The two modes in the third ratio contain different
transition form factors, but the SU(3) symmetry breaking effects in the total
decay amplitudes still cancel to some extent. Hence, the second equality provides
useful information, and is worth consideration.

The data $B(D^0\to \pi^-\pi^+)=(1.455\pm 0.024)\times 10^{-3}$, 
$B(D^0\to \pi^0\pi^0)=(8.26\pm 0.25)\times 10^{-4}$, $B(D^0\to \pi^-K^+)=(1.50\pm 0.07)\times 10^{-4}$ 
and $B(D^0\to \pi^0K_S)=(1.240\pm 0.022)\times 10^{-2}$ lead to
$1.76\pm 0.06$ for the first ratio and $4.32\pm 0.22$
for the second ratio in Eq.~(\ref{r2}), which differ dramatically. The data
$B(D^0\to \pi^+K^-)=(3.950\pm 0.031)\times 10^{-2}$ yield the third ratio 
$3.19\pm 0.062$, which is closer to the second ratio. Note that the modes 
in the first ratio have more final-state pions than those in the 
second and third ratios do.
The apparent difference between the first ratio and the second and third ratios
further supports that the Glauber effect associated with final-state pions is
necessary for explaining the $D\to PP$ data by differentiating the interference 
patterns between the emission and $W$-exchange amplitudes.

\section{CONCLUSION}

In this paper we have identified the residual Glauber divergences
in the $k_T$ factorization for the nonfactorizable annihilation amplitudes in
two-body hadronic heavy meson decays at NLO level. Radiative corrections to these
amplitudes produce not only ordinary collinear logarithms, which are absorbed
into final-state meson wave functions, but imaginary infrared logarithms, which
demand the introduction of additional nonperturbative inputs. It has been shown that 
the Glauber divergences are factorizable in the region with small parton momenta, 
to which the $k_T$ factorizaton theorem applies. It was observed that the resultant
phase factors for the two LO nonfactorizable annihilation diagrams 
are the same. Therefore, the Glauber gluon effect rotates these amplitudes, 
and modifies their interferences with other amplitudes, such that branching ratios 
and direct CP asymmetries of some two-body hadronic heavy meson decays can be changed.
Besides, this work provides a solid theoretical ground for the FAT parametrization of 
two-body hadronic $D$ meson decays proposed a decade ago \cite{Li:2012cfa}.

Because the Glauber phase factors are of perturbative origin, there is no a priori
knowledge about their importance in heavy flavor decays. We have demonstrated that 
the data suggest prominent Glauber effects in pion-involving modes, which have been 
argued for by means of the simultaneous role of a pion as a $q\bar q$ bound state and
a pseudo Nambu-Goldstone boson. The Glauber effects in two-body hadronic $B$ meson decays 
were crucial for resolving the $B\to\pi\pi$ and $\pi K$ puzzles. 
The similar observation has been made here for two-body hadronic
$D$ meson decays into pions: 
the obvious difference among the measured $D\to\pi\pi$, $\pi K$ and $KK$ branching ratios 
reflects the Glauber effects associated with the final-state pions, which differentiate the 
interference patterns between the emission and annihilation ($W$-exchange) amplitudes
in the above modes. We have explained that the Glauber divergences 
are absent in the collinear factorization for heavy meson decays.
A derivation of the Glauber phases in the $k_T$ factorization by
nonperturbative methods for various mesons will help
verifying the mechanism elaborated in this work.

\begin{acknowledgments}

I thank X.D. Gao, X. Liu, S. Mishima, Y.H. Tsai, F.S. Yu and X. Yu for stimulating discussions. 
This work was supported in part by MOST of R.O.C. under Grant No. MOST-107-2119-M-001-035-MY3.

\end{acknowledgments}

\end{document}